\documentclass{webofc}
\usepackage[varg]{txfonts}   
\def\larsoft{\texttt{LArSoft}\xspace}
\def\icc{\texttt{icc}\xspace}
\def\gcc{\texttt{gcc}\xspace}
\def\mkl{\texttt{MKL}\xspace}
\def\fftw{\texttt{FFTW}\xspace}
\begin{document}
\title{Reconstruction for Liquid Argon TPC Neutrino Detectors Using Parallel Architectures}
%
%

\author{\firstname{Sophie} \lastname{Berkman}\inst{1} \and
        \firstname{Giuseppe} \lastname{Cerati}\inst{1}\fnsep\thanks{\email{cerati@fnal.gov}}
        \firstname{Brian} \lastname{Gravelle}\inst{2} \and
        \firstname{Boyana} \lastname{Norris}\inst{2} \and
        \firstname{Allison} \lastname{Reinsvold Hall}\inst{1} \and
        \firstname{Michael} \lastname{Wang}\inst{1}
}

\institute{Fermi National Accelerator Laboratory, Batavia, IL, USA 60510
\and
           University of Oregon, Eugene, OR, USA 97403
          }

\abstract{%
  Neutrinos are particles that interact rarely, so identifying them requires large detectors which produce lots of data. Processing this data with the computing power available is becoming more difficult as the detectors increase in size to reach their physics goals. In liquid argon time projection chambers (TPCs) the charged particles from neutrino interactions produce ionization electrons which drift in an electric field towards a series of collection wires, and the signal on the wires is used to reconstruct the interaction. The MicroBooNE detector currently collecting data at Fermilab has 8000 wires, and planned future experiments like DUNE will have 100 times more, which means that the time required to reconstruct an event will scale accordingly. Modernization of liquid argon TPC reconstruction code, including vectorization, parallelization and code portability to GPUs, will help to mitigate these challenges. The liquid argon TPC hit finding algorithm within the \larsoft framework used across multiple experiments has been vectorized and parallelized. This increases the speed of the algorithm on the order of ten times within a standalone version on Intel architectures. This new version has been incorporated back into \larsoft so that it can be generally used. These methods will also be applied to other low-level reconstruction algorithms of the wire signals such as the deconvolution. The applications and performance of this modernized liquid argon TPC wire reconstruction will be presented.
}
\maketitle
\section{Introduction}
\label{intro}

The US-based neutrino physics program relies on present and future experiments using the Liquid Argon Time Projection Chamber (LArTPC) technology.
The flagship experiment is the future Deep Underground Neutrino Experiment (DUNE)~\cite{Abi:2018dnh}, which will be made of four cryostat modules of approximately 10~kt liquid argon each.
DUNE will be located at the Sanford mine in South Dakota and will detect neutrinos from the beam produced 1300~km away, at Fermilab.
Among the primary science drivers of DUNE are fundamental questions like: "Is the CP symmetry violated in the leptonic sector?", and "What is the neutrino mass hierarchy?".
Smaller scale LArTPC experiments are already operating or will be turning on in the next months.
The MicroBooNE detector~\cite{Acciarri:2016smi} has a mass $m$ of about 170~t, is located at a distance $L$ of about 400~m from the source of the Booster Neutrino Beam (BNB) at Fermilab, and has been taking data since 2015.
Along the same beam line, ICARUS~\cite{Amerio:2004ze} ($m$=760~t, $L$=500~m) and SBND~\cite{Admas:2013xka} ($m$=260~t, $L$=100~m) are currently completing the installation and construction phase, respectively.
These experiments are part of the Short Baseline Neutrino (SBN) program~\cite{Antonello:2015lea}, whose goals are to shed light on the low-energy anomalies reported by LSND~\cite{Aguilar:2001ty} and MiniBooNE~\cite{Aguilar-Arevalo:2018gpe}, possibly compatible with the presence of sterile neutrinos, and to precisely measure neutrino-argon cross sections which are of utmost importance for DUNE.

In LArTPC detectors charged particles produced by neutrino interactions ionize the argon and produce ionization electrons.  These drift towards the anode planes of the detector in a homogeneous electric field applied to the argon volume. There are typically three anode planes, each composed of an array of sense wires, that measure the ionization charge.
A negative or zero voltage is applied to the first two planes, called induction planes, so that the drifting charge passes through the wires and induces a bipolar signal.
A positive voltage is applied to the last plane, called the collection plane, as the ionization electrons are collected and a unipolar signal is measured.
Wires are oriented at different angles in the three planes to allow for three-dimensional reconstruction of the neutrino interaction.

Several factors make reconstruction in LArTPC experiments a challenge: the neutrino interaction vertex can be anywhere in the large argon volume, many different topologies can arise from the neutrino interaction, non-negligible electronic noise may overlap with the signal on the wire waveforms, and, for surface detectors, a contamination of tracks from cosmic rays overlaps with the neutrino signal.
LArTPC is a relatively ``young'' detector technology, so there is not a single reconstruction approach known to provide the best performance across the board. Reconstruction is a field of active development for these detectors, and different techniques are being explored, from more traditional hit-based reconstruction approaches such as Pandora~\cite{Acciarri:2017hat}, to image-based reconstruction approaches exploiting Deep Learning techniques~\cite{Adams:2018bvi}, or even a tomographic-like paradigm~\cite{Qian:2018qbv}.
However, early parts of the reconstruction chain, such as wire signal processing and hit finding, are more mature, and in many cases they are used in common across different approaches.

All included, the reconstruction time in a state-of-the-art experiment such as MicroBooNE is of the order of minutes per event. This is a challenge, especially given the increasing size of the next generation of experiments such as ICARUS or DUNE, with respective sizes order of 5 and 100 times bigger. Large speedups are needed to be able to promptly process the data for those experiments. The key feature of LArTPC detectors that comes to the rescue is their modular structure (cryostats, TPCs, planes, wires) which opens up the possibility for parallel processing of the reconstruction algorithms. In fact, the trend in computing hardware for the last decade indicates that increases in clock frequency for single processors are practically over. This means that speedups are only possible thanks to increasing vector width (enabling parallelism over data, \emph{vectorization}) and thanks to increasing number of processing units (enabling parallelism over different threads).

This work is part of a project\footnote{HEP Event Reconstruction with Cutting Edge Computing Architectures project supported by the DOE SciDAC program, https://computing.fnal.gov/hepreco-scidac4/} whose mission is to accelerate the event reconstruction in high-energy physics experiments using modern parallel computing architectures.
The project typically operates as follows:
we identify key algorithms for the physics outcomes of high energy physics experiments that are also dominant contributors to their reconstruction workflows,
we characterize and re-design those algorithms to make efficient usage of parallelism, both at data- and instruction-level,
we deploy the new code in the experiments' frameworks, and finally we explore their execution on different architectures and platforms.
This paper describes the work performed towards the optimization of the hit finding and signal processing algorithms for LArTPC experiments.

\section{Optimization of Hit Finding Algorithm}

The MicroBooNE detector is composed of about 8k wires which are read out at a frequency of 2 MHz.
Therefore, each wire outputs a binned waveform, with one ADC reading per sampling.
After signal processing (see Sec.~\ref{sec:sp}), regions of interest (ROI) along the wire waveforms are identified.
The ionization signal in a ROI is typically a Gaussian-shaped pulse.
Hit finding is the process to identify such pulses and determine their peak position, width, and amplitude; for more details on the algorithm, see~\cite{Baller:2017ugz}.
While this is a simple algorithm, it can take a significant fraction of the reconstruction workflow time, ranging from a few percent to a few tens of percent of the total
depending on the experiment.
For this algorithm, wires can be independently processed, so it is particularly suitable to demonstrate the speedup potential from the parallelization of LArTPC reconstruction.

We replicated the central version of the hit finder algorithm contained in the \larsoft~\cite{Snider:2017wjd}
repository and used by all Fermilab LArTPC experiments in a standalone code for testing and optimization.
The main difference between the standalone and central versions is that we replaced the Gaussian fit routines based on \texttt{Minuit}+\texttt{ROOT}\xspace\cite{Hatlo:2005,Brun:1997} with a local implementation of the
Levenberg-Marquardt (LM) minimization~\cite{Bevington:1992}.
The LM approach is controlled by a \emph{damping parameter} $\lambda$ so that, when the fit parameters are far from the minimum,
$\lambda$ is large and the algorithm resembles a gradient descent while, when the fit is close to convergence, $\lambda$ becomes smaller and the algorithms approximates
the Gauss-Newton method based on the Hessian minimization.
The fit parameters can be restricted to vary only within user-defined boundaries for better fit stability.
Early tests, before optimizations and without any vectorization or multi-threading, showed that our standalone implementation is about 8 times faster than the original.

\begin{figure*}
\centering
\includegraphics[width=12cm,clip]{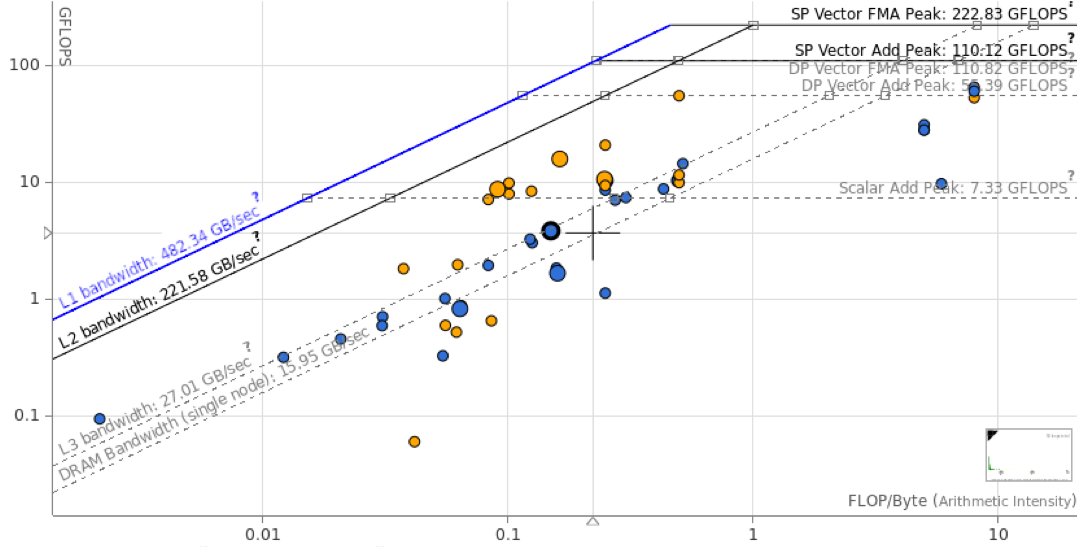}
\caption{Example of a roofline analysis plot obtained with Intel Advisor; the y axis corresponds to the number of floating points per second (in GFLOPS units) and the x axis is the arithmetic intensity (i.e. how many operations are made per byte loaded in the memory registers, in FLOP/Byte units); each dot corresponds to a different function or loop in the code, the size of the dot indicates the exclusive time spent in that function or loop, while the color indicates whether the loop is scalar (blue) or vectorized (orange).}
\label{fig-roofl}
\end{figure*}

\begin{figure*}
\centering
\includegraphics[width=6cm,clip]{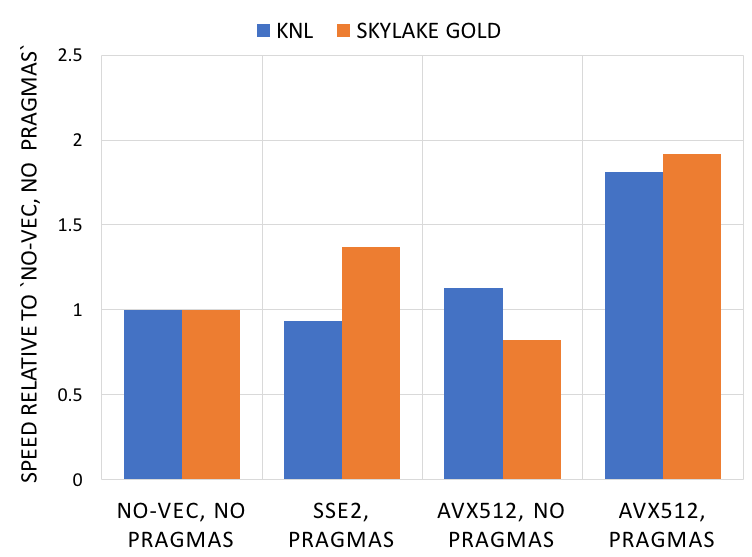}
\caption{Speedups on SKL and KNL for different compilation configurations, defined in terms of instruction sets, or whether compiler directives (\emph{pragma}) are used to instruct the compiler on which loops to vectorize.}
\label{fig-mrqdt-vec}
\end{figure*}

We profiled the computational performance of the standalone code with methods such as the roofline~\cite{Williams:rfl} analysis (Fig.~\ref{fig-roofl}), and
results showed that, even with the LM approach, most of the time is still spent in the minimization algorithm.
The number of iterations needed by the fit to converge varies for each hit candidate so it is difficult to achieve a good performance by vectorizing across multiple hit candidates.
Therefore, we chose to vectorize the most time consuming loops within the LM algorithm, which typically compute the Gaussian parameters or derivatives across waveform data bins. The main limitations in this approach are that only a subset of the code is vectorized, and that the number of bins in an ROI is of the same order as the vector unit size so the relative size of the ``remainder'' part of the loop 
($N\%v$, where $N$ is the number of data bins and $v$ is the vector size) is non negligible.
Despite these limitations, our tests show that both on Intel Xeon Gold 6148 (Skylake, \emph{SKL}) and Intel Xeon Phi Processor 7230 (Knights Landing, \emph{KNL}) processors,
when compiling with \icc 19.0.5.281 and with the AVX-512 instruction set, we achieve close to a factor 2 speedup (Fig.~\ref{fig-mrqdt-vec}).
We also compared our code with the vectorized Trust-Region minimization algorithm included in the Intel Math Kernel Library (\mkl)~\cite{MKL} and found that our fitter is significantly faster.

\begin{figure*}
\centering
\includegraphics[width=6cm,clip]{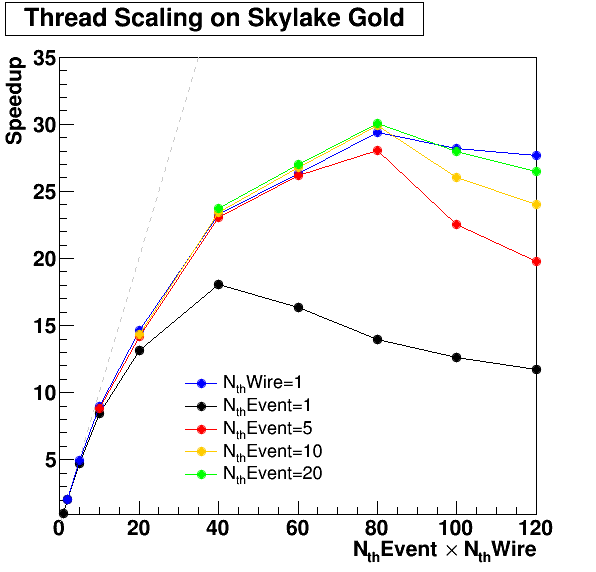}
\hfill
\includegraphics[width=6cm,clip]{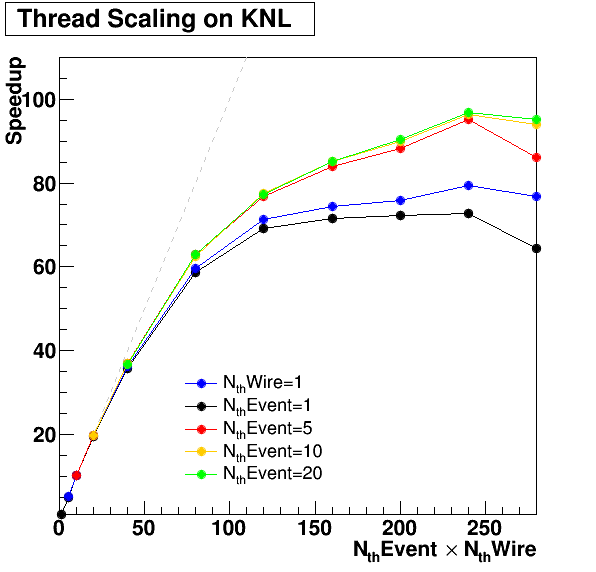}
\caption{Multi-threading speedups with respect to serial execution on SKL (left) and KNL (right). Each curve indicates a different configuration in terms of number of threads, where each configuration has a fixed number of threads per event or per ROI.}
\label{fig-mrqdt-mthr}
\end{figure*}

We implemented multi-threading in the standalone fitting code using \texttt{OpenMP}~\cite{OpenMP}.
The best performance is achieved when using dynamic scheduling and a two-level nested parallelization: over different events using a simple \emph{parallel for}
and over ROIs on wires using a parallel region with \emph{omp for} and a \emph{critical} region for the output synchronization.
Results show that we achieve near ideal scaling at low thread counts and that the speedup increases up to a factor of 30 (95) for 80 (240) threads on SKL (KNL),
see Fig.~\ref{fig-mrqdt-mthr}.

We validated the physics output of our algorithm against the original one by performing a one-to-one comparison of the resulting fit parameters for the same hits, and we observe very little difference between the two, with 98\% of the reconstructed hit times within 0.02 (in waveform bin units) of the original result. 
Similarly to the original, the efficiency of our algorithm is very high across all planes both in MicroBooNE and ICARUS.
It is important to note that very different noise configuration are simulated in these detectors, so our algorithm is robust against large variations in signal to noise ratio, although defining appropriate fit boundaries plays an important role with increasing noise.



We integrated the LM minimization algorithm in \larsoft, and it is available as a plugin for experiments' use.
In this first \larsoft version, the algorithm is compiled with \gcc by default, which limits the vectorization performance, and the multi-threading at wire level is being implemented using \texttt{TBB} because \texttt{OpenMP} is not supported in \larsoft, and the event-level multi-threading is implemented as a feature of the framework.
Testing the \larsoft LM-based hit finder in MicroBooNE and ICARUS reconstruction workflows with a single-threaded execution shows speedups of 12x and 7x, respectively.
In previous data sample production campaigns, the hit finder algorithm in ICARUS took about 40\% of the total reconstruction time, so the experiment will use the LM-based hit finder for its next campaigns.

\section{Studies Towards Optimization of Signal Processing}
\label{sec:sp}

After demonstrating the potential for speedups in LArTPC reconstruction from parallel architectures with the simple hit finder algorithm, we recently focused
on the signal processing algorithm (\emph{SP}), which is the most time consuming part of LArTPC reconstruction.
SP is composed of two steps: first, noise filtering~\cite{Adams:2018dra} is applied to remove excess noise from the waveform signal, and then a deconvolution algorithm~\cite{Adams:2018gbi,Acciarri:2017sde} is applied to correct for the field response and the electronic shaping, thus restoring a Gaussian-shaped pulse.
The time needed to perform the SP scales with the number of wires, so it will grow linearly with the size of detectors; this is a concern for ICARUS, and especially for DUNE.

\begin{figure*}
\centering
\includegraphics[width=6cm,clip]{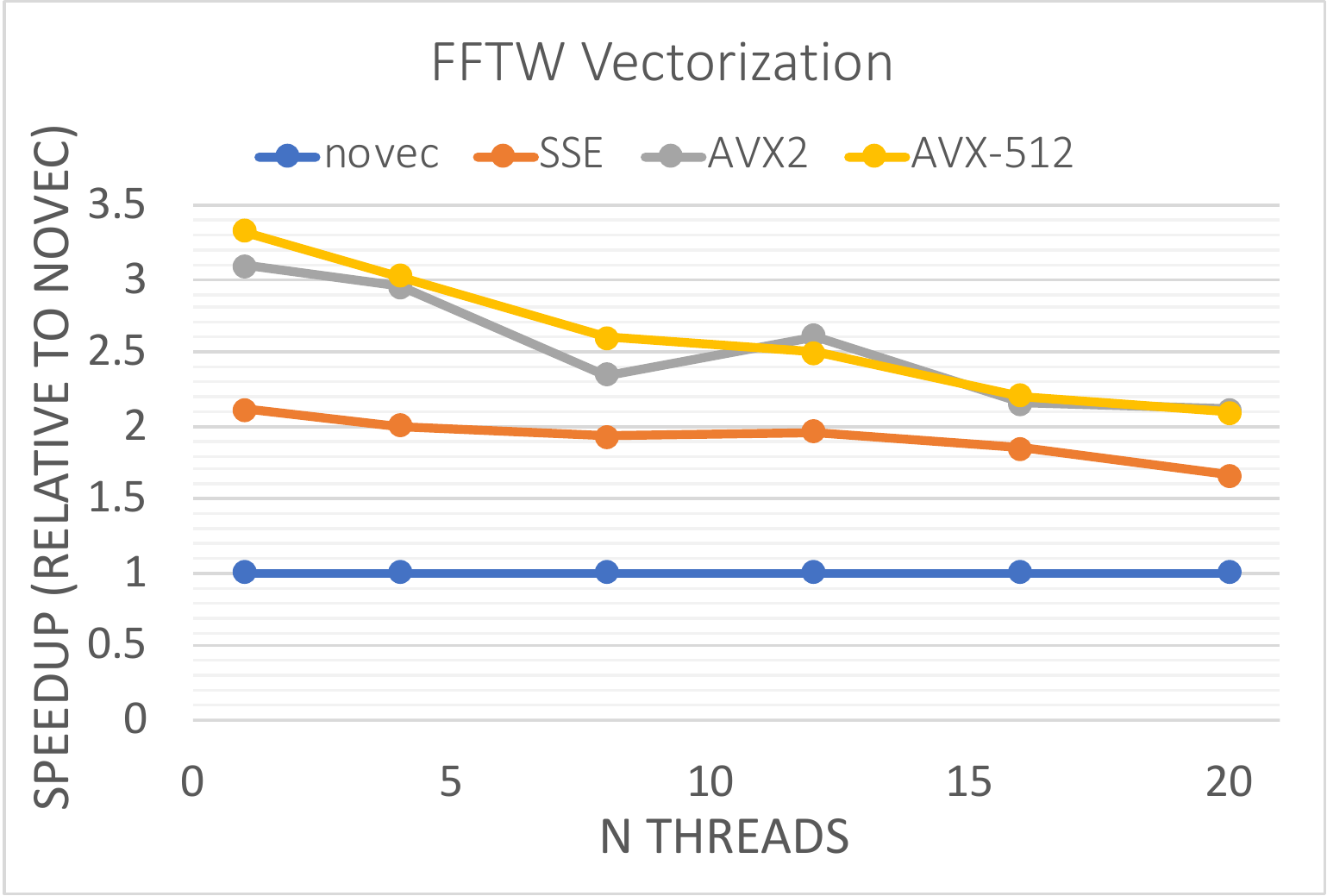}
\hfill
\includegraphics[width=6cm,clip]{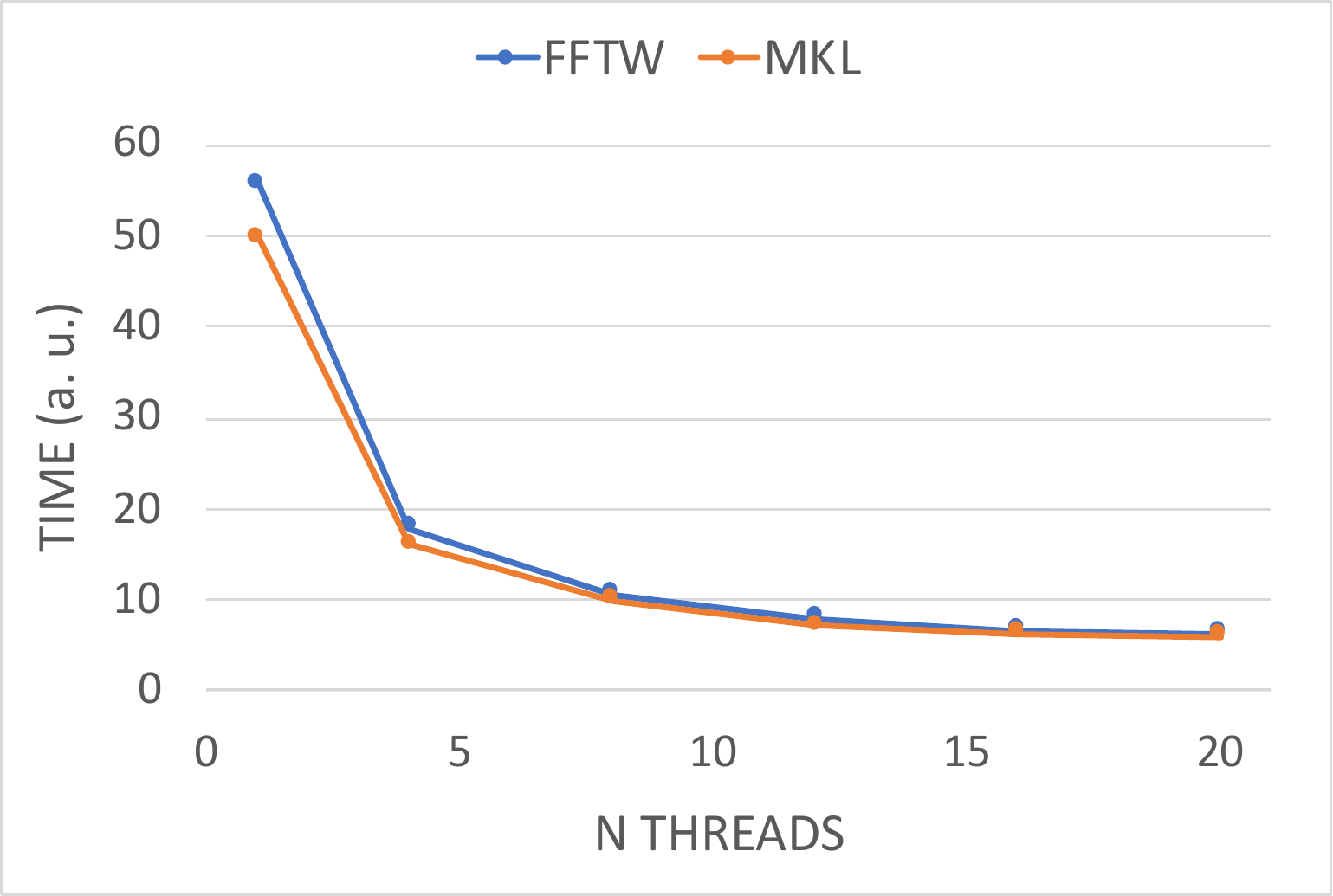}
\caption{Left: speedup of \fftw compiled with different vector instruction sets relative to  compilation without vector instructions (labeled "novec"), as a function of the number of threads.
Right: execution time using \mkl and \fftw libraries as a function of the number of threads.}
\label{fig-fft}
\end{figure*}

We are collaborating with the \larsoft team on the SP algorithms so that, while they are working to make them thread safe, we focus on the core component of these algorithms:
the Fast Fourier Transform (FFT).
The SP algorithms currently in \larsoft are using the FFT implementation from the \fftw library~\cite{Frigo:2005}, compiled with \gcc 7.3.0 using the `core2' x86 option.
We tested \fftw with a more recent version of the compiler (\gcc 9.0.1) and different vector instruction sets: SSE, AVX2, AVX-512.
Results on SKL for a single forward and inverse FFT on wire waveform data are shown in Fig.~\ref{fig-fft}, left.
A speedup of almost a factor of 3.5 is obtained moving from the non-vectorized version to AVX2 or AVX-512, although
possibly due to frequency scaling, the speedup decreases with larger number of threads but is always well above 2.
We also compared our AVX-512 build of \fftw with the \mkl library, which is distributed after compilation with AVX-512.
Figure~\ref{fig-fft}, right shows that there is only marginal difference between the two libraries, with \mkl about 10\% faster than \fftw.
These initial tests currently show thread scaling that is not ideal, at about a speed up of 10 times at 20 threads), so we are working towards identifying and removing unnecessary serial code from within the parallel loops.

\section{Conclusions and Next Steps}

Our project is actively working towards optimizing LArTPC reconstruction algorithms for execution on parallel architectures.
We completed the optimization of the hit finding algorithm, and achieved large speedups thanks to a new minimization algorithm, vectorization, and multi-threading. The code has been ported back to \larsoft, so that experiments are ready to use it in their upcoming production campaigns.
This is the first example of a vectorized and multi-threaded algorithm for LArTPC neutrino experiments.
Work has started to speed up the signal processing algorithms, and our initial studies show that using FFT libraries compiled with AVX-512 will lead to significant improvements.
The next steps for the project will include looking into GPU-compatible implementations for the algorithms and tests at High Performance Computing centers.

\section{Acknowledgements}

This material is based upon work supported by the U.S. Department of Energy, Office of Science, Office of Advanced Scientific Computing Research and Office of High Energy Physics, Scientific Discovery through Advanced Computing (SciDAC) program.

%
%
%

\end{document}